\documentclass[showpacs,preprintnumbers,amsmath,amssymb,12pt]{revtex4}

\usepackage{graphicx}
\usepackage{dcolumn}
\usepackage{bm}
        
\begin{document}
\draft

{\baselineskip0pt
\leftline{\large\baselineskip16pt\sl\vbox to0pt{\hbox{\it Department of
Mathematics and Physics}
               \hbox{\it Osaka City  University}\vss}}

\rightline{\large\baselineskip16pt\rm\vbox to20pt{\hbox{OCU-PHYS-227}
            \hbox{AP-GR-23}
\vss}}%
}
\vskip3cm

\title{Hoop Conjecture in Five-dimensions\\-Violation of
Cosmic Censorship-}

\author{Chul-Moon Yoo\footnote{E-mail:c\_m\_yoo@sci.osaka-cu.ac.jp}, 
Ken-ichi Nakao\footnote{E-mail:knakao@sci.osaka-cu.ac.jp}}
\address{
Department of Mathematics and Physics, Graduate School of Science,
Osaka City University, Osaka 558-8585, Japan
}
\author{Daisuke Ida\footnote{E-mail:daisuke.ida@gakushuin.ac.jp}}
\address{Department of Physics, Gakushuin University, Tokyo 171-8588, Japan
}

\begin{abstract}
We study the condition 
of black hole formation in five-dimensional space-time. 
We analytically solve the constraint equations of five-dimensional 
Einstein equations for momentarily static and conformally flat 
initial data of a spheroidal mass. We numerically search for an apparent horizon 
in various initial hypersurfaces and find both necessary and sufficient 
conditions for the horizon formation in terms of inequalities 
relating a geometric quantity and a mass defined in an appropriate 
manner. In the case of infinitely thin spheroid, 
our results suggest a possibility of naked singularity formation 
by the spindle gravitational collapse in five-dimensional space-time. 
\end{abstract}


\maketitle

\section{Introduction}\label{sec:1}
In an attempt to unify fundamental forces including gravity,
the possibility that the space-time dimensions of our universe is 
higher 
than four has been much discussed. 
Such higher-dimensional theories need mechanism to reduce 
the space-time dimensions down to four,
for example via Kaluza-Klein type compactifications of extra dimensions,
so as to be consistent with the observed world.
The brane world scenario is 
another attractive idea of dimensional reduction.
In this scenario, the standard model particles are confined to the
boundary of a higher-dimensional space-time and only 
gravity can propagate in the extra dimensions. 
Models of the brane world scenario with large extra dimensions compared to
the four-dimensional Planck scale ($\approx 1.6\times 10^{-33}$cm) 
have been considered in some recent works~\cite{BRANE}. 
According to these models, 
the fundamental (namely, higher-dimensional) Planck scale 
may be set to rather low energy scale, even to $1$TeV,
of which low energy effects just alter the short distance behaviour of 
classical gravitational interactions. 
The discrepancy in the gravitational interaction 
between the four and higher-dimensional theories 
arises only at the length 
scale below $0.1{\rm mm}$ so that it is consistent with the 
gravitational experiments~\cite{experiments}. 
In such TeV gravity models, it is suggested that small black holes 
are produced in accelerators, such as the CERN Large Hadron 
Collider~\cite{LHC} or in high energy cosmic ray events~\cite{Feng:2001ib}.
 
In order to understand physical phenomena caused by 
strong gravitational fields, the criterion for
black hole formation is very crucial.
In the case of four-dimensional Einstein gravity, 
such a criterion is well known 
as the hoop conjecture~\cite{HC}.
Hoop conjecture claims that the necessary and sufficient condition
for black hole formation is given by the following;
{\it Black holes with horizons form when and only when a mass
$M$ gets compacted into a region whose circumference in every
direction is ${\cal C}\lesssim 4\pi G_4M$}, 
where $G_4$ is the gravitational constant in four-dimensional 
theory of gravity.
It is remarkable that no serious counterexample against hoop
conjecture has been presented.
However, at first glance, hoop conjecture is not valid 
in higher-dimensional Einstein gravity~\cite{Nakao:2001pn}; 
there is black string solutions in five or higher-dimensions,
which have infinitely long event horizons, while hoop conjecture
claims that any length scale characterizing black hole should be
less than the gravitational length scale determined by the 
Schwarzschild radius. 

Recently, two of the present authors, DI and KN, proposed 
a higher-dimensional version of hoop conjecture~\cite{IN}. 
Here we call it the hyperhoop conjecture in the sense that it is
a possible generalization of the original hoop conjecture;
{\it 
Black holes with horizons form when and only when a mass
$M$ gets compacted into a region whose $(D-3)$-dimensional 
area $V_{D-3}$ in every direction is }
\begin{equation}
V_{D-3}\lesssim G_D M,   \label{eq:hyperhoop}
\end{equation}
where $G_D$ is the gravitational constant in $D$-dimensional 
theory of gravity, and the ($D-3$)-dimensional  
area means the volume of $(D-3)$-dimensional closed submanifold of 
a spacelike hypersurface. 
Hereafter we call this 
$(D-3)$-dimensional closed submanifold {\em the hyperhoop}.
The necessity of the condition~(\ref{eq:hyperhoop}) was
confirmed in the case of momentarily static and 
conformally flat initial data sets 
of an axially symmetric line, disk and thin ring source 
for the five-dimensional Einstein equations~\cite{IN} 
and for the system of 
point-particles~\cite{Yoshino:2002tx}. 
Consistent results with the previous ones were
obtained by  Barrab\'es {\em et al}~\cite{BFL}.
They derived two inequalities for ($D-3$)-dimensional volume as the 
necessary and sufficient conditions for apparent horizon formation 
in the case of a ($D-2$)-dimensional convex thin shell 
collapsing with the speed of light in a $D$-dimensional space-time.

The purpose of the present paper is to study both  
the necessity and 
in particular {\it sufficiency} of the 
inequality~(\ref{eq:hyperhoop}) 
for the horizon formation in different situations from 
the case treated in Ref.~\cite{BFL}. 
We consider the momentarily static and conformally flat 
four-dimensional initial hypersurfaces in which 
a four-dimensional homogeneous spheroid is put
as a gravitational source. This procedure has been implemented 
by 
Nakamura {\em et al.}~\cite{Nakamura:1988zq}.
We apply their method to higher-dimensional case. 
Then, we analytically solve the constraint equations for
five-dimensional Einstein equations.
In order to investigate the validity of hyperhoop conjecture,  
we numerically search for an apparent horizon 
and calculate the ratio $V_2/G_5M$ 
for substantially various hyperhoops.

This paper is organized as follows. In Sec.~\ref{sec:2}, 
assuming five-dimensional Einstein gravity, we derive 
the constraint equations for conformally flat 
initial hypersurfaces and then give analytic solutions 
of these equations for a homogeneous mass of a spheroidal shape. 
In Sec.~\ref{sec:3}, we search for 
an apparent horizon in initial hypersurfaces with various 
shapes of a homogeneous spheroid including infinitely thin case by 
numerically solving a second order ordinary differential equation. 
This equation corresponds to the minimum 
volume condition for a three-dimensional closed submanifold of an 
initial hypersurface. 
The suggestion of the naked singularity formation is given in this 
section. In Sec.~\ref{sec:4}, 
we define $V_2/G_5M$ in a reasonable manner and then 
give a procedure to select the hyperhoop with minimal value 
of $V_2/G_5M$. In Sec.~\ref{sec:5}, we show numerical results 
and  their implication 
to the necessary and sufficient 
condition for the horizon formation. 
Finally, Sec.~\ref{sec:6} 
is devoted to summary.  
In Appendix~\ref{sct:new}, we 
derive analytic solutions for the Newtonian 
gravitational potential of an ellipsoid in arbitrary 
space dimension.
In Appendix~\ref{sct:nec}, the necessary condition of black hole formation 
based on Ref.\cite{IN} is derived.

In this paper, we adopt the unit of $c=1$. We basically follow the 
notations and sign conventions in Ref.\cite{Wald}. 

\section{A Momentarily static spheroid in five-dimensional space-time}
\label{sec:2}

Let us consider an initial data set 
$\left(h_{ab}, K_{ab}\right)$ in 
a four-dimensional spacelike hypersurface $\Sigma$, 
where $h_{ab}$ is the induced metric in $\Sigma$ and 
$K_{ab}$ is the extrinsic curvature 
which represents how $\Sigma$ is embedded in the 
five-dimensional space-time. Denoting the unit normal vector 
to $\Sigma$ by $n^a$, $h_{ab}$ and $K_{ab}$ 
are, respectively, written as 
\begin{eqnarray}
h_{ab}&=&g_{ab}+n_a n_b, \\
K_{ab}&=&-h_a^c\nabla_c n_b,
\end{eqnarray}
where $\nabla_c$ is the covariant derivative in the five-dimensional 
space-time.

The initial data set 
$(h_{ab},K_{ab})$ has to satisfy the Hamiltonian and momentum 
constraints given by 
\begin{equation}
{\cal R}-K_{ab}K^{ab}+K^2=24\pi^2 G_5 \rho
\end{equation}
and
\begin{equation}
D_b\left(K^{ab}-h^{ab}K\right)=12\pi^2G_5 J^a,
\end{equation}
where $\rho$ and $J^a$ are the energy density and energy flux 
for normal line observers to $\Sigma$, $D_a$ and ${\cal R}$ are the 
covariant derivative within $\Sigma$ and the scalar curvature of 
$h_{ab}$, and $G_5$ is the gravitational constant in five-dimensional 
theory of gravity.
In this paper, we focus on momentarily static and conformally 
flat initial hypersurfaces:
\begin{eqnarray}
K_{ab}&=&0\\
h_{ab}&=&f^2\delta_{ab},
\end{eqnarray}
where $\delta_{ab}$ is the metric tensor of four-dimensional 
Euclidean space. 
We also require the axial symmetry in the sense that the metric on
$\Sigma$ have the form
\begin{equation}
dl^2=f^2(R,z)\left[dR^2
    +R^2\left(d\vartheta^2+\sin^2\vartheta d\varphi^2\right)+dz^2\right],
\end{equation}
where $0\leq R<+\infty$ and $-\infty<z<+\infty$ while 
$\vartheta$ and $\varphi$ are the round coordinates. 
Then the momentum constraint leads to zero flux condition $J^a=0$, 
and the Hamiltonian constraint becomes
\begin{equation}
{\partial^2 f \over \partial R^2}
+{2\over R}{\partial f \over\partial R}
+{\partial^2 f \over \partial z^2}
=-4\pi^2 G_5 f^3\rho.\label{eq:rap}
\end{equation}
Here we note that the Hamiltonian constraint~(\ref{eq:rap}) 
is equivalent to the Poisson equation for axi-symmetric Newtonian gravitational potential.
Let us consider the density profile respecting the axial symmetry 
given by
\begin{equation}
f^3\rho=
\left\{\begin{array}{lll}
       2M/\pi^2 a^3b & \mbox{for} & 
       R^2/a^2+z^2/b^2{\leq}1, \\
                ~~~~0 & \mbox{for} & \mbox{elsewhere},
       \end{array}\right. \label{eq:rhon}
\end{equation}
where $a$, $b$  and $M$ are constant parameters. 

We consider the gravitational field of an isolated body, so that we
assume the asymptotic condition given by
\begin{equation}
f\rightarrow 1 ~~~{\rm for}~~r\rightarrow\infty,
\end{equation}
where 
\begin{equation}
r=\sqrt{R^2+z^2}.
\end{equation}
The regular solution is then obtained as 
\begin{eqnarray}
f&=&1-
\frac{4G_5M\left[b(2a+b)R^2+3a^2 z^2 -3a^2b(a+b)
      \right] }{3a^3b{\left( a + b \right) }^2} \nonumber \\
      &&~~~~~~~~~~~~~~~~~~~~~~~~~~~~~~~~~~
{\rm for}~~ \frac{R^2}{a^2}+\frac{z^2}{b^2}~{\leq}~1,
\\
&& \nonumber \\
f&=&1 - \frac{4G_5M}{3 e^4b^4}
\Biggl[
 2R^2-6z^2+3e^2b^2+\sqrt{F^2-e^4b^4} \nonumber \\
&\times&
 \biggl(
  {2e^2b^2 R^2 \over (F-e^2b^2)^2}
  -{2R^2-3z^2+3e^2b^2 \over F-e^2b^2}
  +{3z^2 \over F+e^2b^2}
 \biggr)
\Biggr]\nonumber \\
&&~~~~~~~~~~~~~~~~~~~~~~~~~~~~~~~~~~{\rm for}~~\frac{R^2}{a^2}+\frac{z^2}{b^2}>1,%
\label{eq:outerf}
\end{eqnarray}
where $e$ is the eccentricity defined by
\begin{equation}
e=\sqrt{1-{a^2\over b^2}}
\end{equation}
and $F$ is a function of $R$ and $z$ defined by
\begin{eqnarray}
F&=&F(R,z;a,b)\nonumber \\
 &=&R^2+z^2+\sqrt{4b^2R^2 - 4a^2\left( b^2 - z^2 \right)
 +{\left( a^2 + b^2 - R^2 - z^2 \right)}^2}.\nonumber \\
\end{eqnarray}
\noindent
The detailed derivation of the above solution is given in 
Appendix~\ref{sct:new}. Newtonian gravitational potential 
of an ellipsoid in Euclidean space of arbitrary dimensions is shown 
there. 
Here we only investigate the prolate case $a<b$.

\begin{figure}[htbp]
\begin{center}
\includegraphics[scale=0.85]{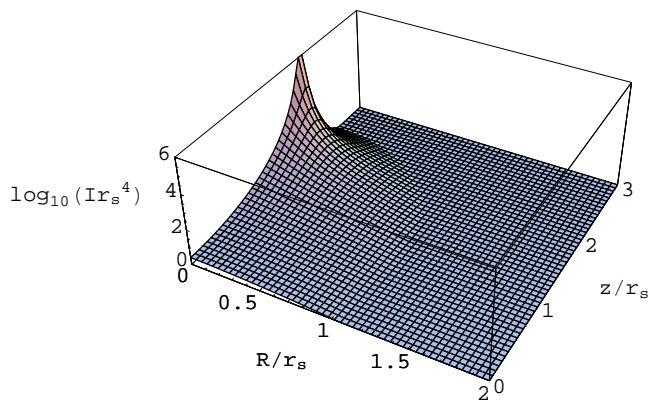}
\caption{The logarithm to the base 10 of the Kretschmann invariant $I$ 
is plotted as function of the coordinates $R$ and $z$ in the case $a=0$ 
and $b=2r_{\rm s}$.}
\label{fig:inv}
\end{center}
\end{figure}

\begin{figure}[htbp]
\begin{center}
\includegraphics[scale=0.85]{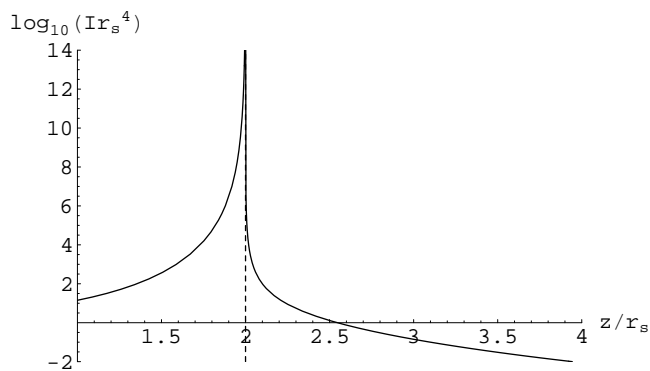}
\caption{The logarithm to the base 10 of the Kretschmann invariant $I$ on 
the polar axis $(R=0)$ is plotted 
as function of $z$ in the case $a=0$ and $b=2r_{\rm s}$.
The value of $I$ is diverge at only $(0,b)$.
}
\label{fig:invR0}
\end{center}
\end{figure}

In the thin limit $a\rightarrow0$ with $M$ and $b$ fixed, 
two disconnected singularities appear at the poles $(R,z)=(0,\pm b)$ 
of the resultant ``singular spheroid''. 
In order to see this, we evaluate the Kretschmann invariant
\begin{equation}
I={\cal R}^{abcd}{\cal R}_{abcd},\label{eq:inv}
\end{equation}
where ${\cal R}_{abcd}$ is the four-dimensional 
Riemann tensor of the spacelike hypersurface. 
Typical examples are shown in 
Figs.~\ref{fig:inv} and \ref{fig:invR0}. 
The coordinate values and the Kretschmann invariant 
$I$ is normalized by
\begin{equation}
r_{\rm s}:=\sqrt{G_5 M}.
\end{equation} 
It will be found in the next section that 
$r_{\rm s}$ is the coordinate radius of the apparent horizon in 
the case of a point source $a=b=0$. 

Here it should be noted that the Kretschmann invariant $I$ is finite 
between these two singularities on the singular spheroid, 
$R=0$ and $|z|<b$. Further, 
we can see that the energy density $\rho$ is also finite there. 
The conformal factor on the surface of the spheroid is given by
\begin{eqnarray}
f&=&f_{\rm sf}(z)\nonumber \\
&:=&1+{4G_5M \over 3ab^2(a+b)^2}\left[b^2(a+2b)-2(b-a)z^2\right].~~
\end{eqnarray}
Therefore we find that the energy density at the surface 
becomes in the thin limit $a\rightarrow0$ as 
\begin{equation}
\rho=\rho_{\rm sf}(z):={2M\over \pi^2a^3 b f_{\rm sf}^{3}}
\longrightarrow {27b^8 \over 256\pi^2 G_5^3 M^2 (b^2-z^2)^3}.
\end{equation}
Here note that the inequality $f(R,z)\geq f_{\rm sf}(z)$ is satisfied 
within the spheroid and hence $0\leq\rho(R,z)\leq \rho_{\rm sf}(z)$. 
Therefore $\rho$ is finite except 
for the poles $z=\pm b$ even in the thin limit $a\rightarrow0$. 
This fact means that the scalar polynomials of the five-dimensional 
Riemann tensor are also finite if the stress of matter fields is 
assumed to be reasonable. For example, assuming the dust as the matter and
adopting Gaussian normal coordinate, Einstein equations leads to
\begin{equation} 
{\partial K^a{}_b \over \partial t}={\cal R}^a{}_b
-4\pi^2 G_5\rho ~\delta^a{}_b
\end{equation}
on the momentarily static initial hypersurface, where ${\cal R}^a{}_b$ 
is the four-dimensional Ricci tensor of this hypersurface. 
Here note that ${\cal R}^a{}_b$ is finite in the region of 
finite Kretschmann invariant $I$ since the metric of the spacelike 
hypersurface is positive definite. Therefore the finiteness of 
the energy density $\rho$ guarantees that the time derivative 
of the extrinsic curvature $K^a{}_b$ is finite in the region 
of finite Kretschmann invariant $I$.  
This means that the scalar polynomials of Riemann tensor of five-dimensional
space-time are also everywhere finite except for the poles 
of the singular spheroid since these are expressed as polynomials 
of $\partial K^a{}_b/\partial t$, ${\cal R}^a{}_b$, $K^a{}_b$ 
and $D_cK^a{}_b$ in the Gaussian normal coordinate. 
Further, we can see that this 
singular spheroid except for the poles corresponds to spatial infinity. 
Consider a curve $z=\zeta(R)$ connecting two points $R=R_1$ and $R=R_2$ 
($R_1<R_2$) and 
assume that both $R_1$ and $R_2$ are sufficiently small and  
$0<\zeta(R)<z_{\rm max}<b$. %
In this situation, the function $F$ is written as 
\begin{equation}
F(R,\zeta;0,b)=b^2+\frac{2b^2}{b^2-\zeta^2}R^2+\mathcal{O}(R^3).
\end{equation}
Hence substituting this for Eq.~(\ref{eq:outerf}), we find
\begin{equation}
f\left(R,\zeta\right)={8(b^2-\zeta^2)^{3/2}G_5 M \over 3b^4 R}+\mathcal{O}(R^0).
\end{equation}
The proper length between $R=R_1$ and $R_2$ along the curve $z=\zeta(R)$ is 
bounded below as
\begin{eqnarray}
&&\int_{R_1}^{R_2}f(R,\zeta)\sqrt{1+\left({d\zeta\over dR}\right)^2}~dR
\nonumber \\
&&\geq\int_{R_1}^{R_2}f(R,\zeta)dR
\simeq {8G_5 M\over 3b^4}
\int_{R_1}^{R_2}(b^2-\zeta^2)^{3/2}{dR\over R} \nonumber \\
&&\geq{8G_5 M\over 3b^4}(b^2-z_{\rm max}^2)^{3/2}\int_{R_1}^{R_2}{dR\over R}
={8G_5 M\over 3b^4}(b^2-z_{\rm max}^2)^{3/2}\ln{R_2 \over R_1}. 
\end{eqnarray}
We can see from the above equation that the proper length diverges 
in the limit of $R_1\rightarrow 0$ with $R_2$ fixed. 
Therefore each point on the singular spheroid except for the poles, 
$R=0$ and $|z|<b$, is spacelike infinity. 
\section{Apparent horizons}\label{sec:3}

In a momentarily static initial hypersurface in 
five-dimensional asymptotically flat space-time, 
an apparent horizon is a three-dimensional closed marginal surface. 
Because of the axial symmetry of the initial hypersurface, 
the apparent horizon will also be axially symmetric 
and thus will be expressed by $r=r_{\rm m}(\xi)$ 
in the present case, where 
\begin{equation} 
\xi=\arctan{R\over z}.
\end{equation} 
Then $r=r_{\rm m}(\xi)$ is a closed marginal surface only if  
$r_{\rm m}(\xi)$ satisfies 
\begin{eqnarray}
&&{\ddot r}_{\rm m}-{4 {\dot r}_{\rm m}{}^2\over r_{\rm m}}
-3r_{\rm m}+\frac{r_{\rm m}{}^2+{\dot r}_{\rm m}{}^2}{r_{\rm m}}
\nonumber\\&&
\times\left[
\frac{2{\dot r}_{\rm m}}{r_{\rm m}}\cot\xi
-{3\over f}\left({\dot r}_{\rm m}\sin\xi+r_{\rm m}\cos\xi\right)
\frac{\partial f}{\partial z}
+{3\over f}\left({\dot r}_{\rm m}\cos\xi-r_{\rm m}\sin\xi\right)
\frac{\partial f}{\partial R}\right]=0
\label{eq:joubi}
\end{eqnarray}
with boundary conditions 
${\dot r}_{\rm m}=0$ at $\xi=0$ and $\pi$, 
where a dot means the derivative with respect to $\xi$. 
The derivation of 
Eq.~(\ref{eq:joubi}) is shown in Appendix~\ref{sec:margi}.
Since the present system has a reflection symmetry 
with respect to $z=0$, the apparent horizon should satisfy 
\begin{equation}
{\dot r}_{\rm m}=0~~{\rm at}~~{\xi= \frac{\pi}{2}}.
\label{eq:BC}
\end{equation}
In the case of a point source $a=b=0$, 
we can analytically solve Eq.~(\ref{eq:joubi}) and find 
\begin{equation}
r_{\rm m}=r_{\rm s}.
\end{equation} 

\begin{figure}[htbp]
\begin{center}
\includegraphics[scale=0.8]{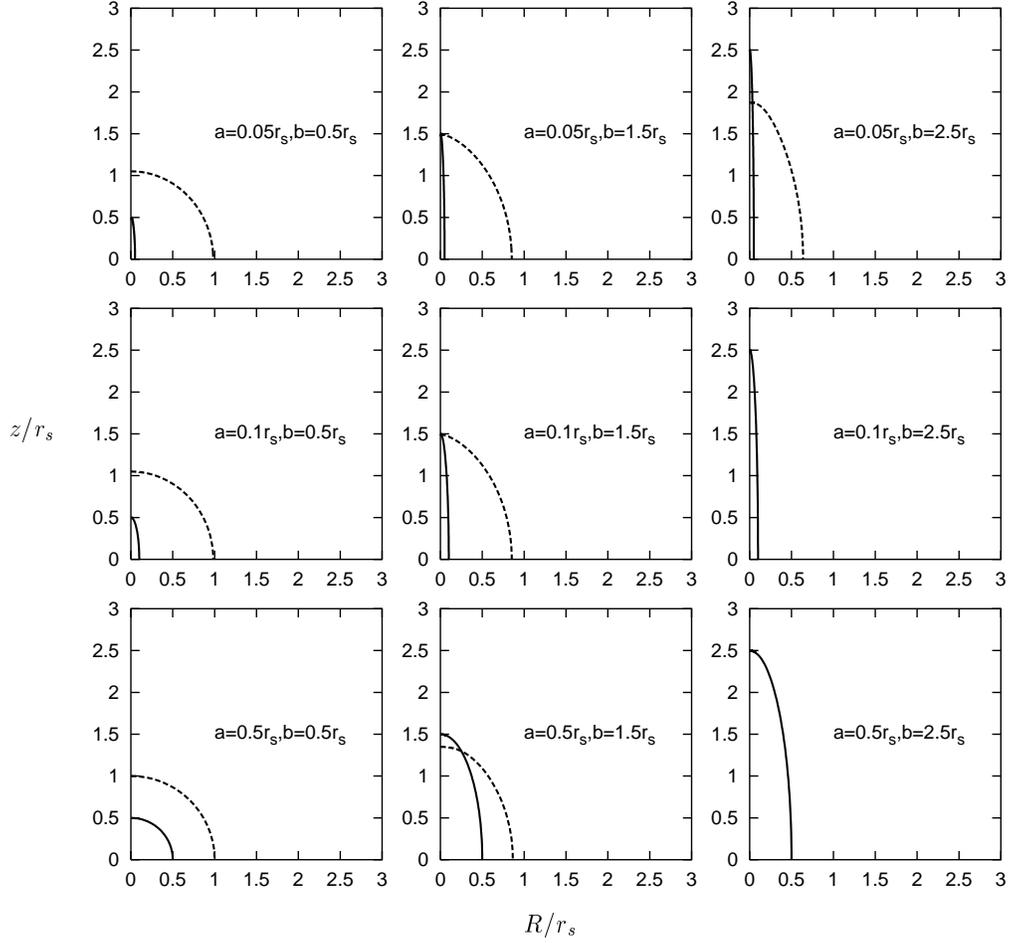}
\caption{Apparent horizons for each shape of the four-dimensional spheroid are 
depicted in $(R,z)$-plane. The solid line shows the surface 
of the spheroid. The dashed line shows the apparent horizon 
if it is present. 
}
\label{fig:aph}
\end{center}
\end{figure}

Replacing derivatives with respect to $\xi$ 
in Eq.~(\ref{eq:joubi}) by finite differences, 
we numerically search for solutions of this equation 
by relaxation method~\cite{ref:SMMN}. 
If apparent horizons exist in the initial hypersurface, 
we can find solutions of Eq.~(\ref{eq:joubi}). 
Typical examples are shown in Fig.~\ref{fig:aph}. 
The coordinate values are normalized by $r_{\rm s}$. 

\begin{figure}[htbp]
\begin{center}
\includegraphics[scale=1.0]{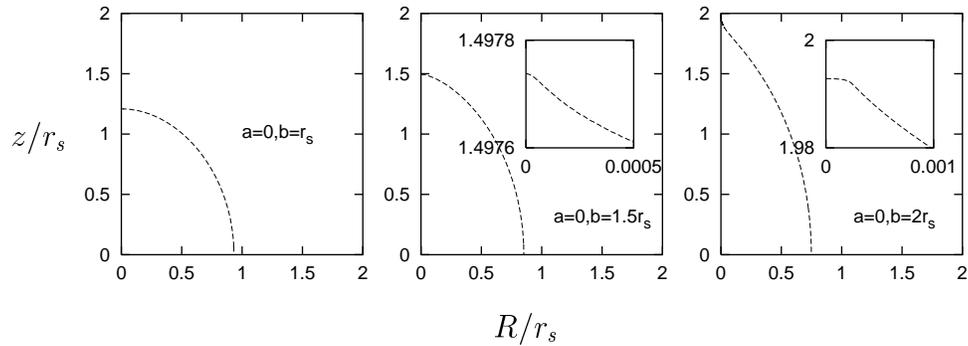}
\caption{In the case of $a=0$, apparent horizons for each $b$ are 
depicted.
}
\label{fig:aphs}
\end{center}
\end{figure}

In the case of singular source $a=0$, we also find solutions of 
Eq.~(\ref{eq:joubi}) satisfying the boundary condition~(\ref{eq:BC}). 
In the case of $b\leq1.48 r_{\rm s}$, there is 
an apparent horizon 
enclosing whole the singular source. By contrast, for 
$b\geq 1.49r_{\rm s}$, the marginal surface covers only a central part 
of the singular source, and the space-time singularities at the poles 
$(R,z)=(0,\pm b)$ are not enclosed by the
marginal surface (see Fig.~\ref{fig:aphs}). 
In this case, this marginal surface is not 
a closed three-surface and thus is not the apparent horizon 
since as mentioned in the previous section, 
$|z|<b$ on the polar axis $R=0$ is the spacelike infinity. 
This result implies that a long enough spindle singular source 
can produce naked singularities, 
which is quite different from the singular line source in Ref.~\cite{IN}.

\section{How to check hyperhoop conjecture}\label{sec:4}

We have presented the statement of hyperhoop conjecture in Sec.~\ref{sec:1}. 
Here we should note that in general, the mass $M$ in hyperhoop 
conjecture (and also in hoop conjecture) is not the 
total mass of the system but 
the mass encircled by the hyperhoop (the hoop). Therefore to check 
the sufficiency of this conjecture, we have to confirm that 
in an initial hypersurface without an apparent horizon, 
there is no hyperhoop satisfying 
\begin{equation}
V_{D-3} \lesssim G_D M_{\rm in}, \label{eq:hutou}
\end{equation}
where $M_{\rm in}$ is not the total mass of the system but the 
one included in the region which is encircled by this hyperhoop. 

Since we cannot try all the possible hyperhoops 
in $\Sigma$, we focus on the axi-symmetric 
hyperhoops which have the reflection symmetry with respect to 
$z=0$. Further, these hyperhoops 
are expressed in the form $r=r_{\rm h}(\xi)$ 
and $\vartheta=\pi/2$ since the spheroid is assumed to be 
prolate $a<b$.
(In the oblate case, we should consider two-surface of 
$z=0$ and $R=$constant, where the hyperhoop is 
parameterized by two parameters $\xi$ and $\varphi$.)
The two-dimensional 
area $V_2$ of the hyperhoop 
$r=r_{\rm h}(\xi)$ and $\vartheta=\pi/2$ is then given by 
\begin{equation}
V_2=\int_0^{2\pi}d\varphi\int^{\pi}_0 d\xi 
f^2\sqrt{{\dot r}_{\rm h}{}^2
+r_{\rm h}{}^2}~r_{\rm h}\sin\xi=4\pi\int^{\pi/2}_0 d\xi 
f^2\sqrt{{\dot r}_{\rm h}{}^2
+r_{\rm h}{}^2}~r_{\rm h}\sin\xi,
\end{equation}
where we have 
taken account of the reflection symmetry in the second equality. 

In the framework of general relativity, there is no 
unique definition of the mass in a quasilocal manner 
although the total mass is well defined for the isolated 
system. This is one of reasons why it is very difficult to 
formulate precisely the hoop and also hyperhoop conjectures.   
However, despite of this mathematical indefiniteness, 
the hoop and hyperhoop conjectures might be useful in 
understanding black hole formation processes on the physical ground. 
In this sense, all the reasonable definitions of quasilocal 
mass will be meaningful in the hoop and hyperhoop conjectures 
since these might give the results not so different 
from each other. Here we adopt the following simple definition 
for the mass $M_{\rm in}$ as 
\begin{equation}
M_{\rm in}=8\pi\int^{\pi/2}_0d\xi
\int^{r_{\rm h}\left(\xi\right)}_0 dr \rho f^3 r^3\sin^2\xi.
\label{eq:Min-definition}
\end{equation}
Then we numerically calculate $V_2/G_5 M_{\rm in}$ for various 
hyperhoops in an initial hypersurface of a spheroid. 
Hereafter for notational simplicity, we introduce 
\begin{equation} 
\Gamma:={V_2\over 16\pi G_5 M_{\rm in}}, \label{eq:gam}
\end{equation}
where $16\pi G_5M_{\rm in}$ is the minimal value of $V_2$ in the 
case of a point source whose mass is $M_{\rm in}$.

Our first task is the selection of 
relevant hyperhoops 
from all the possible hyperhoops expressed 
by $r=r_h(\xi)$ and $\vartheta=\pi/2$ with the reflection symmetry 
with respect to $z=0$. 
A hyperhoop is represented as a continuous curve 
from a point on $z$-axis to a point on $R$-axis in 
the first quadrant of $(R,z)$-plane. 
Thus first, we select a point $z=z_0$ on $z$-axis 
and consider various hyperhoops which start from this point. 
Since the spheroid of the mass is assumed to be prolate, 
we expect that the significant hyperhoops are also prolate, 
and thus we restrict the curves within the region of 
$R^2+z^2\leq z_0^2$. 
The number of possible curves is still infinite. 
Therefore we impose further restrictions. Consider 
100 
points $\left(R,z\right)=\left(R^{(1)}_{j_i}(z_0),z_i(z_0)\right)$ 
$(i,j_i=1,2,..,10)$ within this 
spherical region; $z_i(z_0)$ is determined in the following manner 
\begin{equation}
z_{i}(z_0)={z_0\over10}(10-i),
\end{equation}
and then $R_{j_i}^{(1)}(z_0)$ is given as 
\begin{equation}
R^{(1)}_{j_i}(z_0)={\sqrt{z_0{}^2-z_{i}{}^2(z_0)}\over 10}j_i.
\end{equation}
We consider the curves 
composed of ten straight lines connected at 
$(R,z)=\left(R^{(1)}_{j_i}(z_0),z_i(z_0)\right)$; $i$ is in order  
from $1$ to $10$, and then for each 
$i$, $j_i$ is appropriately chosen among ten integers from 
1 to 10.  

\begin{figure}[htbp]
\begin{center}
\includegraphics[scale=0.75]{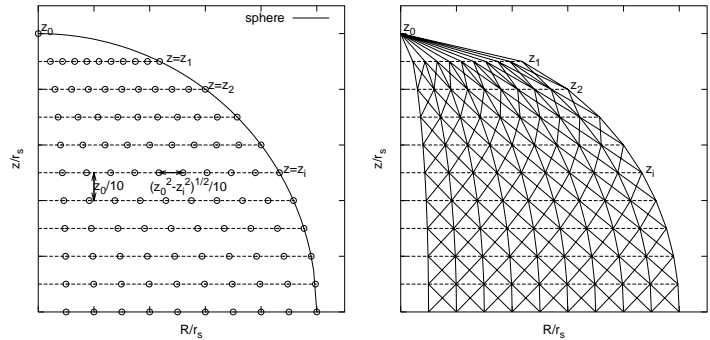}
\caption{
The connecting points (left figure) and 
the hyperhoops which we calculate 
for a value of $z_0$ (right figure) in first search.}
\label{fig:keiro1}
\end{center}
\end{figure}

\begin{figure}[htbp]
\begin{center}
\includegraphics[scale=0.7]{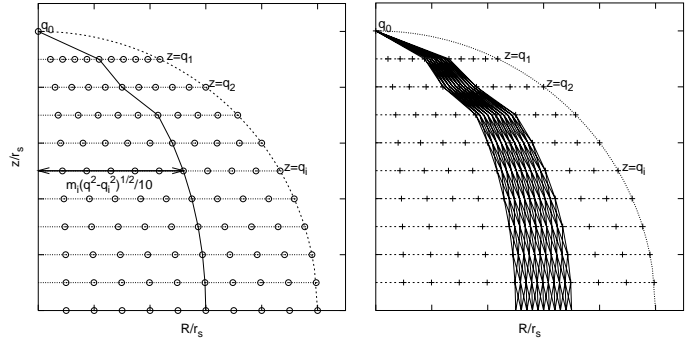}
\caption{
The hyperhoop having the smallest value of $\Gamma$ 
in first search (left figure) and the hyperhoops 
which we calculate for a value of $q_0$ (right figure) in second search 
.}
\label{fig:keiro2}
\end{center}
\end{figure}

By investigating several randomly chosen hyperhoops, we found that 
sharply bended one might have a value of $\Gamma$ 
larger 
than the ones not so sharply bended. 
Hence in the systematic numerical search, a line 
from the point of $(R,z)=\left(R^{(1)}_{j_i}(z_0),z_i(z_0)\right)$ to 
$(R,z)=\left(R^{(1)}_{j_{i+1}}(z_0),z_{i+1}(z_0)\right)$ is adopted as 
a constitutive one 
of the hyperhoop only if $j_{i+1}$ is equal to $j_i$ or  
$j_i\pm 1$ for $j_i>1$ and is equal to $j_i$ or $j_i+1$ for $j_i=1$. 
All the possible constitutive lines are depicted in Fig.~\ref{fig:keiro1}. 
This means that  we consider only the hyperhoops each of which is 
a connected set of ten constitutive lines. 
Then we calculate $\Gamma$ for each 
hyperhoop and search for the minimal one which is specified 
by a set of ten integers $\{m_i\}$ in the manner of $j_i=m_i$. 
Further for several values of $z_0$, we carry out the same calculations 
as the above. 
We select the values of $z_0$ at even intervals $0.1r_s$ 
and search for the hyperhoop with the smallest value of 
$\Gamma$. 
Finally, we obtain the minimal one which is specified by a set of 
ten integers and one real number $\{m_i,q\}$, where $q$ is the value of $z_0$.
The above hyperhoop $\{m_i,q\}$ might not be exactly minimal 
since the hyperhoops obtained by the above procedure are 
too restrictive. Thus we might find hyperhoops smaller than 
the one $\{m_i,q\}$ in the following refinement. 
We consider 
a neighbourhood 
$R^{(1)}_{m_{i}-1}(q_0)\leq R \leq R^{(1)}_{m_{i}+1}(q_0)$
of the hyperhoop $\{m_i,q_0\}$, where $q_0$ is a value which is equal or near to $q$.
In this region, we put further grid points at 
\begin{equation}
R=R^{(2)}_{k_i}(q_0)
=R^{(1)}_{m_{i}-1}(q_0)
+{\sqrt{q_0^2-q_i^2(q_0)}\over 50}(k_i-1),~~
(k_i=1,2,\cdots,11),
\end{equation}
where
\begin{equation}
q_i(q_0)=\frac{q_0}{10}\times (10-i).
\end{equation}
Then by the same procedure as in the previous search for 
the minimal hyperhoop, we will obtain the hyperhoop with 
$\Gamma$ smaller than the previous one (see Fig.~\ref{fig:keiro2}). 
Further for several value of $q_0$ in the vicinity of $q$, 
we carry out the same calculations as the above. 
We select the 
11 
values 
\begin{equation}
q-0.1r_s+0.02r_s(l-1),~~(l=1,2,\cdots,11)
\end{equation}
as $q_0$
and search for the hyperhoop with the smallest value of 
$\Gamma$. 

\section{Numerical Results}\label{sec:5}

For various $a$ and $b$, numerical results of the minimal value of $\Gamma$ 
are listed in TABLE~\ref{tab:re}. 
Hereafter we denote the minimal value of $\Gamma$ by 
$\Gamma_{\rm min}$ which is a function of $a$ and $b$. 

We see from TABLE~\ref{tab:re} that 
$\Gamma_{\rm min}$ is not smaller than unity 
in the case of no apparent horizon.
This implies that the inequality~(\ref{eq:hutou}) is really a sufficient condition 
in the situations studied here.

\begin{table*}[htbp]
\caption{The minimal values of $\Gamma$ $\left(\Gamma_{\rm min}\right)$ is shown 
for each $a$ and $b$. 
The existence of an apparent horizon is represented 
by (Yes):exist and (No):not exist.}
\label{tab:re}
\begin{center}
\begin{tabular}{|c||c|c|c|c|c|c|c|}
\hline
$b \backslash a$ & 0 & 0.01 & 0.1 & 0.3 & 0.5 & 0.7 & 0.9 \\
\hline
1.0&1.05(Yes)&1.05(Yes)&1.05(Yes)&1.05(Yes)&1.04(Yes)&1.03(Yes)&1.01(Yes) \\
1.5&1.11(No)&1.11(Yes)&1.11(Yes)&1.10(Yes)&1.10(Yes)&1.09(No)&1.08(No) \\
2.0&1.17(No)&1.17(Yes)&1.17(Yes)&1.17(No)&1.16(No)&1.16(No)&1.17(No) \\
2.5&1.23(No)&1.23(Yes)&1.23(No)&1.23(No)&1.23(No)&1.24(No)&1.26(No) \\
3.0&1.29(No)&1.29(Yes)&1.29(No)&1.29(No)&1.30(No)&1.31(No)&1.35(No) \\
3.5&1.35(No)&1.34(No)&1.34(No)&1.34(No)&1.35(No)&1.38(No)&1.43(No) \\
4.0&1.40(No)&1.38(No)&1.38(No)&1.38(No)&1.40(No)&1.44(No)&1.51(No) \\
\hline
\end{tabular}
\end{center}
\end{table*}

Next, let us study the necessity of hyperhoop conjecture~(\ref{eq:hutou}). 
By the investigation of the singular line source of  
a ``constant'' line energy density studied in Ref.~\cite{IN}, 
we obtain a necessary condition as follows 
\begin{equation}
V_2\lesssim\frac{\pi}{2}
\times16\pi G_5M.
   \label{eq:nec}
\end{equation}
The derivation of this quantity $\pi/2$ is shown in Appendix~\ref{sct:nec}.
The counterexample for this condition is not found in TABLE~\ref{tab:re}.
Here, it is again noted that a closed marginal surface is formed 
only when $b\leq1.48r_{\rm s}$, and thus our result of $a=0$ suggests the 
formation of naked singularities in the case $b\geq 1.49r_{\rm s}$. 
If it is true, the naked singularities might form by the gravitational collapse
starting from the initial data of nonvanishing but sufficiently small $a$ 
and $b$ larger than $1.49r_{\rm s}$. 
If a naked singularity exists, the apparent horizon 
does not necessarily mean the existence of a black hole. 
Therefore, although the apparent horizon forms 
for  $a\in [0.01,0.5]$ and $b=1.5r_{\rm s}$ (see TABLE~\ref{tab:re}), 
these results do not necessarily mean the black hole formation. 

Finally, in the present situation, we find the following inequalities, 
\begin{equation}
{\rm Necessary ~condition:}~~ V_2\lesssim\frac{\pi}{2}
16\pi G_5M,
\end{equation}
\begin{equation}
{\rm Sufficient ~condition:}~~V_2\lesssim16\pi G_5M. \label{eq:suf}
\end{equation}
Our numerical results suggest that the necessary condition is 
not identical to the sufficient condition.

\section{Summary}\label{sec:6}
We have investigated the condition of 
apparent horizon formation in the case of momentarily static 
and conformally flat initial data of a spheroidal mass 
in the framework of five-dimensional Einstein gravity. 
All our results are consistent with the hyperhoop conjecture.
Particularly, we confirmed the sufficiency of the
inequality~(\ref{eq:hyperhoop}). 
(More precisely, inequality (\ref{eq:hutou}) holds 
in the present case.)

We also consider the limit of infinitely prolate spheroid.
The gravitational field of such spheroids are singular and have a nontrivial structure.
The poles at the both ends of this 
singular spheroid are the space-time singularities since the
Kretschmann invariant diverges at the poles of the spheroid. 
On the other hand, both the Kretschmann invariant and the energy density 
are finite elsewhere.
 
We find that the singular spheroid is spacelike infinity except for the poles. 
Furthermore, when the singular spheroid is sufficiently long, in an appropriate sense,
no apparent horizon appears.
This property can be regarded as being peculiar to nonuniform distribution
of material energy, because a uniform line energy density is always enclosed 
by an apparent horizon in spite of its length~\cite{IN}. 

One might wonder if the singular spheroid is a 
counterexample to the hyperhoop conjecture, since it is infinitely thin 
and hence there seems to be a hyperhoop satisfying the inequality~(\ref{eq:hyperhoop}). 
However as we have shown, the singular spheroid is 
not a counterexample to the hyperhoop conjecture. 
In order to understand its reason, we have to note two important 
features of the initial hypersurface 
studied here. First, the proper area $V_2$ of a hyperhoop tightly 
encircling the surface of the spheroid is not necessarily 
smaller than those encircling outside of the spheroid, 
since the conformal factor in the inner region takes the value 
larger than that in the outer region. 
Second, the proper area $V_2$ of a hyperhoop tightly encircling 
the spheroid does not necessarily become smaller when the coordinate 
size of the spheroid characterized by $a$ and $b$ becomes smaller, 
since the conformal factor in the spheroid of the smaller size becomes 
larger if the mass $M$ is fixed. 

The difference between 
the present case and the previous work~\cite{IN} is that 
the line energy density vanishes continuously at the poles 
in the present case, while it vanishes discontinuously 
at the poles  
in the previous case.
In general, if an infinitely thin line object forms 
by the gravitational collapse, it might have a line 
energy density which continuously vanishes at the end of the matter 
distribution. Therefore, the naked singularity formation 
seems to be generic in the axi-symmetric gravitational collapse 
of highly elongated matter distribution in five-dimensional space-time, 
although we would need numerical simulation to have a definite evidence
for the naked singularity formation~\cite{Shapiro:1990}.
This might strongly depend on the spacetime 
dimension~\cite{Patil:2003yp} 
and this is also a future work. 

\section*{Acknowledgements}
We are grateful to H.~Ishihara and colleagues in the astrophysics and 
gravity group of Osaka City University for helpful discussion and 
criticism.  This work is supported by the Grant-in-Aid for Scientific 
Research (No.16540264) from JSPS. 

\appendix
\section{Newtonian potentials of a homogeneous ellipsoid 
in $D$-dimensional space-time}\label{sct:new}
In this section, we extend Newtonian potentials of a homogeneous ellipsoid 
in four-dimensional space-time to $(n+1)$-dimensional space-time. 
The reader may refer to Ref.~\cite{EFE} about the potentials of four-dimensional 
space-time.

We want to obtain the potentials of the homogeneous ellipsoid of which 
bounding ellipsoid is 
\begin{equation}
\sum^n_{i=1}\frac{x_i^2}{a_i^2}=1.\label{eq:source}
\end{equation}
At the beginning of derivation, we define the 
{\em {\textquotedblleft}homoeoid''}. 
\\\\
{\em Definition}. A $n$-dimensional homoeoid is a shell bounded 
by two similar concentric $n$-dimensional ellipsoids 
in which strata of equal density are also $n$-dimensional ellipsoids 
that are concentric with and similar to the bounding ellipsoids.
\\\\
Following theorem and corollary is derived in the same way 
as the case of four-dimensional space-time~\cite{EFE}.
\\\\
{\em Theorem}. The potential at internal point of a $n$-dimensional 
homoeoid is constant.
\\\\
{\em Corollary}. The equipotential surfaces external 
to a thin $n$-dimensional homoeoid are $n$-dimensional ellipsoids 
confocal to the homoeoid.
\\\\
We can obtain the potential of a thin $n$-dimensional homoeoid expressed as 
\begin{equation}
\sum^n_{i=1}\frac{x_i^2}{a_i^2}=1,~~(a_1<a_2<\cdots<a_n),\label{eq:ellip}
\end{equation}
using $n$-dimensional ellipsoidal coordinates $(y_1,y_2,\cdots,y_n)$ 
which satisfy
\begin{equation}
\sum^n_{i=1}\frac{x_i^2}{a_i^2-y_j}=1,~~(j=1,2,\cdots,n)\label{eq:renritu}
\end{equation}
and
\begin{equation}
y_1<y_2<\cdots<y_n. 
\end{equation}
We can solve Eq.~(\ref{eq:renritu}) for $x_i^2$ and
\begin{equation}
x_i^2=\frac{\prod_{j=1}^{n}(a_i^2-y_j)}{\prod_{k\neq i}(a_i^2-a_k^2)}. 
\label{eq:ri2}
\end{equation}
Therefore
\begin{equation}
\frac{\partial x_i}{\partial y_j}=\frac{x_i}{2(y_j-a_i^2)}. \label{eq:bib}
\end{equation}
The metric is expressed as 
\begin{equation}
\sum_{i=1}^{n}dx_i^2=\frac{1}{4}\sum_{i=1}^{n}
\frac{\prod_{k\neq i}(y_i-y_k)}{\prod_{j=1}^{n}(a_j^2-y_i)}dy_i^2, 
\label{eq:rsenso}
\end{equation}
thus
\begin{equation}
\det g=\left(\frac{1}{4}\right)^n\frac{\prod_{j=1}^{n}
\left[\prod_{k\neq j}(y_j-y_k)\right]}{\prod_{i,j}(a_i^2-y_j)}.
\label{eq:detg}
\end{equation}

In order to obtain the potential $\Phi_N$ of a thin $n$-dimensional homoeoid, 
we only have to solve following equation because of the theorem 
and the corollary.
\begin{equation}
\triangle\Phi_N(y_1)=0\label{eq:tri}
\end{equation}
with
\begin{equation}
\Phi_N \rightarrow
-\frac{M}{(n-2)r^{n-2}}~~~{\rm for}~~~r\rightarrow\infty, \label{eq:bcth}
\end{equation}
where $M$ is constant which correspond to the mass of the thin $n$-dimensional homoeoid and $r$ is the length from the center of the homoeoid.
It can be seen that $-y_1\sim r^2$ at infinity.
Using ellipsoidal coordinates to Eq.~(\ref{eq:tri}), we obtain
\begin{equation}
\frac{\partial}{\partial y_1}\left[\prod_{i=1}^n\sqrt{a_i^2-y_1}
\frac{\partial\Phi_N(y_1)}{\partial y_1}\right]=0. \label{eq:lap0p}
\end{equation}
The solution of Eq.~(\ref{eq:lap0p}) with (\ref{eq:bcth}) is 
\begin{equation}
\Phi_N(u)=-\frac{n-2}{2}M\int^{\infty}_{u}\frac{1}{\prod_{i=1}^n
\sqrt{a_i^2+u}}du, \label{eq:homopo}
\end{equation}
where $u=-y_1$.

Integrating this potentials of thin homoeoid which is foliated 
in all region of ellipsoid in the same manner 
as four-dimensional space-time~\cite{EFE}, 
we can finally obtain the Newtonian potential of a $D$-dimensional 
homogeneous ellipsoid.
The integration can be done as follows.
We can express the thin homoeoids which is similar and concentric to the 
ellipsoid~(\ref{eq:source}) as
\begin{equation}
\sum^n_{i=1}\frac{x_i^2}{a_i^2}=m^2,
\end{equation}
where $m$ is constant and we assume $0\leq m \leq1$.
Let us consider the homogeneous homoeoid bounded by two ellipsoids 
$\sum^n_{i=1}\frac{x_i^2}{a_i^2}=m^2$ and
$\sum^n_{i=1}\frac{x_i^2}{a_i^2}=(m+dm)^2$, where $dm$ is small deviation. 
The mass of this homoeoid is 
\begin{equation}
\omega_na_1a_2\cdots a_n\rho m^{n-1}dm,\label{eq:homomass}
\end{equation}
where $\omega_n$ and $\rho$ is $n$-dimensional solid angle and the density 
of the homoeoid respectively.

First, we derive the potential in outer region of homogeneous ellipsoid.
Substituting (\ref{eq:homomass}) into $M$ of (\ref{eq:homopo}), we can 
obtain the potential of the
homoeoidal element~(\ref{eq:homomass}) at $(x'_1,x'_2,\cdots,x'_n)$
\begin{equation}
-\frac{n-2}{2}\omega_na_1a_2\cdots a_n\rho m^{n-1}dm
\int^\infty_{\lambda(m^2)}\frac{du'}{\prod^n_{i=1}\sqrt{a_i^2m^2+u'}},
\label{eq:homopote}
\end{equation}
where $\lambda$ is the largest root of $\sum^n_{i=1}\frac{{x'_i}^2}{m^2a_i^2
+\lambda}=1$.
Integrating this equation about $0\leq m\leq 1$, we have
\begin{eqnarray}
-\frac{n-2}{2}\omega_na_1a_2\cdots a_n\rho \int^1_0dmm^{n-1} 
\int^\infty_{\lambda(m^2)}\frac{du'}{\prod^n_{i=1}\sqrt{a_i^2+u'}}
&&\nonumber \\
=-\frac{n-2}{4}\omega_na_1a_2\cdots a_n\rho \int^1_0dm^2
\int^\infty_{\mu(m^2)}\frac{du}{\prod^n_{i=1}\sqrt{a_i^2+u}},&&\nonumber\\
\end{eqnarray}
where $u$ and $\mu$ defined by $u'=m^2u$ and $\lambda(m^2)=m^2\mu(m^2)$ 
respectively.
Now, we can invert the order of integrations because $\mu(m^2)$ is the 
monotone decreasing function of $m^2$. 
Since $\mu \rightarrow \infty$ when $m \rightarrow 0$ and $\mu=\lambda$ 
when $m=1$, 
we can write
\begin{equation}
\Phi_N=-\frac{n-2}{4}\omega_na_1\cdots a_n\rho
\int^\infty_{\lambda(1)}\frac{(1-m^2(u))}{\prod^n_{i=1}\sqrt{a_i^2+u}}du
\label{eq:sotofpote}
\end{equation}
in outer region, where 
\begin{equation}
m^2(u)=\sum^n_{i=1}\frac{x_i^2}{a_i^2+u}.
\end{equation}

Next, we derive the potential in inner region of homogeneous ellipsoid 
at the point 
$(x'_1,x'_2,\cdots,x'_n)$ which satisfy 
$\sum^n_{i=1}\frac{x'^2_i}{a_i^2}=m'^2$, 
where $m'$ is constant and $0\leq m'<1$.
On the one hand, the potential of homogeneous ellipsoid bounded by 
$\sum^n_{i=1}\frac{x^2_i}{a_i^2m'^2}=1$ at 
$(x'_1,x'_2,\cdots,x'_n)$ is 
\begin{equation}
-\frac{n-2}{4}\omega_na_1\cdots a_n \rho 
\int^\infty_0\frac{(m'^2-m^2(u))}{\prod^n_{i=1}\sqrt{a_i^2+u}}du,
\label{eq:elipote}
\end{equation}
where we use (\ref{eq:sotofpote}).
On the other hand, the potential of homogeneous homoeoid bounded by 
$\sum^n_{i=1}\frac{x^2_i}{a_i^2}=m'^2$ and$\sum^n_{i=1}\frac{x^2_i}{a_i^2}=1$ 
at $(x'_1,x'_2,\cdots,x'_n)$ is obtained by integration of 
(\ref{eq:homopote}) from $m=m'$ to $m=1$, and we have
\begin{equation}
-\frac{n-2}{4}\omega_na_1a_2\cdots a_n \rho (1-m'^2)
\int^\infty_0\frac{du}{\prod^n_{i=1}\sqrt{a_i^2+u}},\label{eq:homo}
\end{equation}
Adding together (\ref{eq:homo}) and (\ref{eq:elipote}), we can obtain 
the required potential
\begin{eqnarray}
\Phi_N=-\frac{n-2}{4}\omega_na_1\cdots a_n\rho
\int^\infty_0\frac{(1-m^2(u))}{\prod^n_{i=1}\sqrt{a_i^2+u}}du~~~~
\end{eqnarray}
in inner region.

If we have same radiuses to the directions corresponding to coordinates 
$x_i,x_{i+1},\cdots$ in (\ref{eq:source}), we can introduce 
multipolar coordinates to these and calculate as same.

\section{the necessary condition of black hole formation 
in five-dimensional space-time}\label{sct:nec}

We consider the singular line source studied in Ref.~\cite{IN}. 
The energy density is given by  
\begin{equation}
f^3\rho=\frac{1}{4\pi R^2}\frac{G_5M}{2b}\delta(R)\theta(b-|z|)
\label{eq:spindle}
\end{equation}
where $\theta$ is the Heaviside's step function and 
the ``length'' of this line source is given by $2b$. 
In this case, the solution of the Hamiltonian constraint~(\ref{eq:rap}) 
is given by 
\begin{equation}
f=1+\frac{G_5M}{2bR}\left(\arctan\frac{z+b}{R}-\arctan\frac{z-b}{R}\right).
\end{equation}
As shown in Ref.~\cite{IN}, this line source is always covered 
by an apparent horizon.

In order to obtain the necessary condition of an apparent horizon formation, 
we have to calculate the values of $\Gamma_{\rm min}$ defined 
in Sec.~\ref{sec:5}. 
Here, we consider the hyperhoops which are expressed in the 
form $r=r_h(\xi)$ and $\vartheta=\pi/2$ with the same symmetry 
as discussed in Sec.~\ref{sec:4} (axi-symmetry and reflection 
symmetry with respect to $z=0$). 

In this case, the hyperhoop which intersects the line source 
has infinite two-dimensional area $V_2$. 
In order to see this, consider a hyperhoop expressed as 
$\vartheta=\pi/2$ and $z=\eta(R)$, where $|\eta(0)|<b$ 
so that the hyperhoop intersects the line source. 
We focus on a segment $R_1<R<R_2$ and $z>0$ of this hyperhoop. 
We assume that $R_1$ and $R_2$ are sufficiently small 
and $0<\eta(R)<b$ on this segment. 
The conformal factor is approximately given by
\begin{equation}
f\simeq \frac{\pi \left(2\eta-b\right)G_5M}{4bR\eta} 
\end{equation}
on this segment of the hyperhoop.
Hence, the area of this segment is bounded below as 
\begin{eqnarray}
&&\int^{R_2}_{R_1}\int^{2\pi}_{0}f(R,\eta)^2R
\sqrt{1+\left(\frac{d\eta}{dR}\right)^2}d\varphi dR
\nonumber \\ 
&&\geq2\pi\int^{R_2}_{R_1}f(R,\eta)^2RdR\simeq\frac{\pi^3 G_5^2M^2}{8b^2}\int^{R_2}_{R_1}
\left(2-\frac{b}{\eta}\right)^2\frac{dR}{R}\nonumber \\ 
&&\geq\frac{\pi^3 G_5^2M^2}{8b^2}\int^{R_2}_{R_1}\frac{dR}{R} 
=\frac{\pi^3 G_5^2M^2}{8b^2}\ln \frac{R_2}{R_1}.
\end{eqnarray}
We can see from the above equation that the area of the hyperhoop 
diverges in the limit of $R_1\rightarrow0$ with $R_2$ fixed. 
Therefore we have to consider only the hyperhoop which entirely 
encircle the line source, and hence we find 
\begin{equation}
\Gamma_{\rm min}=\left(\frac{V_2}{16\pi G_5M_{\rm in}}\right)_{\rm min}
=\frac{\left(V_2\right)_{\rm min}}{16\pi G_5M}, 
\end{equation}
where $\left(V_2\right)_{\rm min}$ is the area of the hyperhoop  
which entirely encircle the source and has the smallest area. 
In order to evaluate $\left(V_2\right)_{\rm min}$, 
we focus on the hyperhoop $r=r_{\rm a}(\xi)$ and $\vartheta=\pi /2$ 
which satisfy following minimum area condition 
\begin{equation}
\delta V_2=0, \label{eq:vari}
\end{equation}
where $\delta V_2$ is the small variation of $V_2$ for slight 
deformation of the hyperhoop 
which keeps it on $\vartheta=\pi/2$ and the symmetry holds.
Namely, Eq.~(\ref{eq:vari}) leads to the Euler-Lagrange equation 
for the Lagrangian $L=V_2\left(r_{\rm a},\dot{r_{\rm a}}\right)$ as 
\begin{eqnarray}
&&{\ddot r}_{\rm a}-{3 {\dot r}_{\rm a}{}^2\over r_{\rm a}}
-2r_{\rm a} 
+\frac{r_{\rm a}{}^2+{\dot r}_{\rm a}{}^2}{r_{\rm a}} 
\nonumber \\
&&\times\left[
\frac{{\dot r}_{\rm a}}{r_{\rm a}}\cot\xi
-{2\over f}\left({\dot r}_{\rm a}\sin\xi+r_{\rm a}\cos\xi\right)
\frac{\partial f}{\partial z}
+{2\over f}\left({\dot r}_{\rm a}\cos\xi-r_{\rm a}\sin\xi\right)
\frac{\partial f}{\partial R}\right]=0.\label{eq:deforma}
\end{eqnarray}
We impose following boundary conditions 
so that every part of the hyperhoop locally satisfy Eq.~(\ref{eq:deforma})
\begin{equation}
{\dot r}_{\rm a}=0~~~~~{\rm at}~~\xi=0,~\frac{\pi}{2}.\label{eq:bcforma}
\end{equation}
Then $r=r_{\rm a}(\xi)$ is the hyperhoop of the minimum area if and only if 
$r_{\rm a}(\xi)$ satisfies Eq.~(\ref{eq:deforma}) 
with the boundary condition~(\ref{eq:bcforma}).
We adopt the area of this hyperhoop as $(V_2)_{\rm min}$.

\begin{figure}[htbp]
\begin{center}
\includegraphics[scale=0.7]{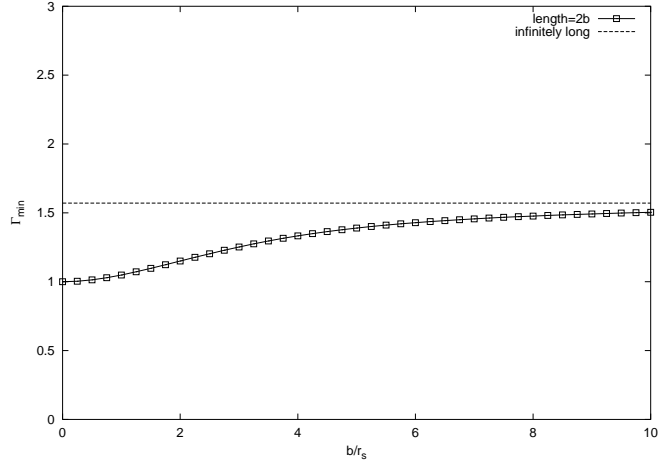}
\caption{
The value of $\Gamma_{\rm min}$ is plotted as a 
function of $b$. 
The dashed line bound this value above.
This line is the corresponding quantity for the infinitely long spindle source 
which have line density $M/2b$. 
}
\label{fig:deb}
\end{center}
\end{figure}

We numerically search for solutions of Eq.~(\ref{eq:deforma}) 
with (\ref{eq:bcforma}). 
Accordingly, the solutions always can be found and the hyperhoop 
always encircle the source in spite of its length.
The value of $\Gamma_{\rm min}$ is depicted in Fig.~\ref{fig:deb} as 
a function of $b$.

The value of $\Gamma_{\rm min}$ monotonically increases with 
$b$ but has a finite limit for $b\rightarrow \infty$, while  
an apparent horizon always covers this line source. 
Therefore it is necessary 
for apparent horizon formation that $\Gamma_{\rm min}$ is smaller 
than this asymptotic value. 
The asymptotic value
will be obtained by evaluating the corresponding quantity 
of the infinitely long source case. 
Let us consider the infinitely long singular line source 
whose density profile is given by 
\begin{equation}
f^3\rho=\frac{1}{4\pi R^2}\frac{G_5M}{2b}\delta(R).
\end{equation}
In this case, we can easily solve the Hamiltonian constraint~(\ref{eq:rap}) 
and obtain 
\begin{equation}
f=1+\frac{\pi G_5M}{2bR}.
\end{equation}
The area $V_{\rm c}$ of the cylindrical two-surface $R=R_0$ and 
$\vartheta=\pi/2$ with coordinate length $2b$ is 
\begin{equation}
V_{\rm c}=2\pi Rf|_{R=R_0} \times 2bf|_{R=R_0},
\end{equation}
where $2\pi Rf|_{R=R_0}$ is the proper length of circle 
around singular line source and $2bf|_{R=R_0}$ 
is the proper length of the cylinder measured 
along the $z$-direction. 
The minimal value of $V_{\rm c}$  is realized when  
$R_0=\pi G_5M/2b$, and in its value is equal to $8\pi^2 G_5M$. 
This minimal value might be almost equal to 
$\left(V_2\right)_{\rm min}$ of the singular line 
source~(\ref{eq:spindle}) if $b$ is much longer than $r_{\rm s}$. 
As a result, the asymptotic value of $\Gamma_{\rm min}$ for 
$b\rightarrow\infty$ with the mass $M$ fixed is evaluated as 
$\pi/2$. 
\section{The derivation of the equation for a marginal surface}\label{sec:margi}

In this section, we show the derivation of Eq.(\ref{eq:joubi}). 
Here, we generalize Ref.\cite{ref:SMMN} to the $D$-dimensional case. 

We denote the spacelike unit vector outward from and normal to 
the marginal surface by $s_a$, and the spacelike unit vectors 
spanning the marginal surface are denoted by $(e^A)_a$, 
where $A=1,..,D-2$. All these vectors are chosen to be orthogonal 
to each other and to the unit vector normal to the 
initial hypersurface $n_a$. Then the future directed outward null 
vector $l^a$ orthogonal to the marginal surface is written by 
\begin{equation}
l^a=n^a+s^a.
\end{equation}
The expansion $\chi$ of this null vector is defined by 
\begin{equation}
\chi=\delta_{AB}(e^A)^a(e^B)^b\nabla_bl_a=(h^{ab}-s^as^b)(K_{ab}-D_bs_a)
,\label{eq:chi}
\end{equation}
where $h_{ab}$ and $K_{ab}$ are the induced metric and the extrinsic 
curvature defined in Sec.\ref{sec:2} respectively. 
The marginal surface is a closed $(D-2)$-dimensional spacelike submanifold 
such that the outward null vector orthogonal to the 
$(D-2)$-dimensional spacelike submanifold has vanishing expansion. 
Hence, the equation to define the marginal surface is given by 
$\chi=0$. In the momentarily static case, this equation reduces to 
\begin{equation}
\delta_{AB}(e^A)^aD_a(e^B)^bs_b=0.\label{eq:momestaj}
\end{equation}

In the situation presented in this paper, coordinates of points on 
the marginal surface are represented as 
\begin{equation}
x^\mu=
\left(
r_{\rm m}(\xi)\cos\xi,r_{\rm m}(\xi)\sin\xi,\vartheta,\varphi
\right) 
\end{equation}
and following vectors are tangent to the marginal surface as 
\begin{eqnarray}
\frac{\partial x^\mu}{\partial\xi}
&=&\left(\dot{r}_{\rm m}\cos\xi-r\sin\xi,
\dot{r}_{\rm m}\sin\xi+r_{\rm m}\cos\xi,0,0\right),~~~~~\\
\frac{\partial x^\mu}{\partial\vartheta}&=&(0,0,1,0),\\
\frac{\partial x^\mu}{\partial\varphi}&=&(0,0,0,1).
\end{eqnarray}
Hence, we can obtain the components of $s^a$ and $(e^A)^a$ as
\begin{eqnarray}
s^\mu&=&\frac{1}{f\sqrt{r_{\rm m}^2+\dot{r}_{\rm m}^2}}
\left(\dot{r}_{\rm m}\sin\xi+r_{\rm m}\cos\xi,-\dot{r}_{\rm m}\cos\xi+r_{\rm m}\sin\xi,0,0\right),\label{eq:vectors4} \\
(e^1)^\mu&
=&\frac{1}{f\sqrt{r_{\rm m}^2+\dot{r}_{\rm m}^2}}
\left(\dot{r}_{\rm m}\cos\xi-r_{\rm m}\sin\xi,\dot{r}_{\rm m}\sin\xi+r_{\rm m}\cos\xi,0,0\right),\label{eq:vectors1}\\
(e^2)^\mu&=&\frac{1}{fR}(0,0,1,0), \label{eq:vectors2}\\
(e^3)^\mu&=&\frac{1}{fR\sin\vartheta}(0,0,0,1).\label{eq:vectors3}
\end{eqnarray}
Substituting Eqs.(\ref{eq:vectors4})$-$(\ref{eq:vectors3}) 
into Eq.(\ref{eq:momestaj}), we obtain Eq.(\ref{eq:joubi}).

\end{document}